\documentclass[prb,preprintnumbers,twocolumn,showpacs,nobalancelastpage]{revtex4}

\usepackage{amsmath,amssymb,amsfonts}
\usepackage{bm,graphicx,color}

\newcommand{\convz}{\ast}

\newcommand{\ret}{{\text{r}}}
\newcommand{\adv}{{\text{a}}}

\newcommand{\les}{<}

\newcommand{\PES}{\mathcal{I}}

\newcommand{\flavorf}{\uparrow}
\newcommand{\flavorc}{\downarrow}
\newcommand{\Dloc}{Q}
\newcommand{\dloc}{q}
\newcommand{\Bloc}{R}
\newcommand{\bloc}{r}

\newcommand{\cG}{\mathcal{G}}

\newcommand{\z}{{+}}
\newcommand{\e}{{-}}
\newcommand{\ze}{{\z\e}}
\newcommand{\ez}{{\e\z}}
\newcommand{\ee}{{\e\e}}
\newcommand{\zz}{{\z\z}}
\newcommand{\ft}{\tilde}

\newcommand{\CC}{\mathcal{C}}

\newcommand{\TC}{\text{T}_{\CC}}
\newcommand{\DC}[1]{\partial^\CC_{#1}}

\newcommand{\delC}{\delta^\CC}

\newcommand{\tb}{\bar t}
\newcommand{\tmin}{t_{\text{min}}}
\newcommand{\tmax}{t_{\text{max}}}

\newcommand{\Ks}{{\Vk\sigma}}

\newcommand{\vect}[1]{{\bm #1}}
\newcommand{\Vk}{{\vect{k}}}

\newcommand{\VR}{{\vect{R}}}
\newcommand{\Vq}{{\vect{q}}}

\newcommand{\ekin}{{E}}
\newcommand{\ke}{{\hat{\vect{k}}_e}}
\newcommand{\pp}{{|\hspace{-1pt}|}}

\newcommand{\expval}[1]{\langle{#1}\rangle}

\newcommand{\myast}{{{}\hspace*{-0.1em}\ast\hspace*{-0.1em}{}}}
\newcommand{\mydagger}{{\dagger}}
\newcommand{\phdagger}{{\phantom{\mydagger}\!}}
\newcommand{\ETAL}{{\em et al.}}

\makeatletter\def\Dated@name{}\makeatother

\begin{document}

  \title{Measuring correlated electron dynamics with
    time-resolved photoemission spectroscopy}

  \author{Martin Eckstein}

  \author{Marcus Kollar}

  \affiliation{Theoretical Physics III, Center for Electronic
    Correlations and Magnetism, Institute for Physics, University of
    Augsburg, 86135 Augsburg, Germany}

  \date{September 25, 2008}

  \begin{abstract}
    Time-resolved photoemission experiments can reveal fascinating
    quantum dynamics of correlated electrons.  However, the
    thermalization of the electronic system is typically so fast that very
    short probe pulses are necessary to resolve the time evolution of
    the quantum state, and this leads to poor energy resolution due to
    the energy-time uncertainty relation.  Although the photoemission
    intensity can be calculated from the nonequilibrium electronic
    Green functions, the converse procedure is therefore difficult.
    We analyze a hypothetical time-resolved photoemission experiment
    on a correlated electronic system, described by the
    Falicov-Kimball model in dynamical mean-field theory, which
    relaxes between metallic and insulating phases.  We find that the
    real-time Green function which describes the transient behavior
    during the buildup of the metallic state cannot be determined
    directly from the photoemission signal. On the other hand, the
    characteristic collapse-and-revival oscillations of an excited
    Mott insulator can be observed as oscillating weight in the center
    of the Mott gap in the time-dependent photoemission spectrum.
    \end{abstract}


  \maketitle

  \section{Introduction}

  Pump-probe experiments with femtosecond time resolution can record
  various nonequilibrium processes in solids directly in the time
  domain, including those induced by the Coulomb interaction between
  electrons, or the scattering of electrons on defects and phonons.
  In these experiments excitation of the sample and characterization
  of the excited state are accomplished by two distinct laser pulses
  (pump and probe) which hit the sample with controlled time delay,
  and either optical or photoemission spectroscopy may be used as
  probe technique.  The pump-probe setup has been used to investigate
  the dynamics of molecules,\cite{Zewail00}
  semiconductors,\cite{Axt04} and metals\cite{Petek97} for more than
  two decades. More recently, such time-resolved experiments were also
  performed on several strongly correlated materials close to a phase
  transition, where many degrees of freedom contribute to the dynamics
  on very different time scales.%
  \cite{Ogasawara00,Iwai03,Chollet05,Okamoto07,Perfetti06,Kuebler07}
  
  In such nonequilibrium solid-state experiments it is a major 
  challenge to distinguish the electronic dynamics from other degrees 
  of freedom.  This is crucial in particular for the Mott metal-insulator
  transition,\cite{Imada98} which is driven by the Coulomb interaction
  between electrons moving in a crystal lattice. An entirely new
  perspective on this phenomenon would open up if one could observe
  the transition as it happens in real time and, e.g., monitor
  the formation of well-defined quasiparticles as the system goes from
  an insulating to a metallic state. In fact, in several Mott and
  charge-transfer insulators the transition to a metallic state can be
  induced by a laser pump
  pulse.\cite{Ogasawara00,Iwai03,Okamoto07,Perfetti06} So far these
  experiments have focused on the relaxation back to the insulating
  state, which involves coupling to degrees of freedom other than the
  valence band electrons, and is much slower than the buildup of the
  metallic state after the pump pulse. In the simplest case the 
  observed relaxation is described by the two-temperature 
  model,\cite{Allen87} i.e., as the cooling of a hot electron gas which 
  is coupled to the colder
  lattice.\cite{Perfetti06,Freericks08b}
  The true dynamics of the electronic system
  has so far been observed only in simple metals, by looking at the
  thermalization of pump-excited electron distributions due to
  electron-electron scattering.\cite{Fann92,Petek97} In strongly 
  correlated materials, thermalization is apparently much faster.
  Sufficient time resolution is now becoming available due to recent advances 
  in femtosecond laser techniques,\cite{Steinmeyer99,Hentschel01} 
  which have already allowed to investigate some solid-state systems 
  even on the attosecond time scale.\cite{Cavalieri07}

  In equilibrium, electronic properties of correlated materials
  can be obtained directly from conventional photoemission
  spectroscopy with continuous light beams.\cite{pes1,pes2} By contrast, time-resolved 
  measurements are likely to be restricted
  by the frequency-time uncertainty of the probe pulse: The energy $\epsilon$
  of occupied states in the solid from which photoelectrons are
  released is determined from the kinetic energy of the
  photoelectrons, the work function of the solid, and the photon
  energy $E_{\gamma}$ $=$ $\hbar\omega$; when the measurement pulse has
  finite duration $\delta t$, the latter is determined only up to an
  uncertainty $\delta E_{\gamma}$ $\gtrsim$ $\hbar/\delta t$.  In a
  strongly correlated electron system we would expect that typical
  relaxation times are directly related to the energy scales that
  appear in the spectrum, such as the bandwidth or the Mott gap.  
  In this case all information on the initial energy
  $\epsilon$ is lost for pulses which are short enough to resolve the
  electronic dynamics. The equilibrium interpretation of conventional
  photoemission data in terms of the electronic spectrum of the solid
  thus becomes meaningless in this limit.
  
  A full theory for time-resolved photoemission spectroscopy (TRPES), which
  covers both the nonequilibrium effects of the electronic state and
  also the consequences of the frequency uncertainty of the pulse, was
  presented recently by Freericks \ETAL.\cite{Freericks08} Their
  approach extends existing theories of conventional photoemission
  spectroscopy to the case where the sample is not in
  equilibrium and measurement pulses have a finite time duration. The
  photoemission intensity as a function of the probe pulse delay time
  is related to electronic one-particle real-time Green functions of
  the sample,\cite{Freericks08} which fully incorporate the
  nonequilibrium many-body dynamics after the pump pulse. This is
  in contrast to earlier Green function approaches,\cite{Sakaue05}
  which treat pump and probe on the same (perturbative) level.
  The relation to real-time Green functions allows to
  make direct contact to recent progress in nonequilibrium many-body
  theory, such as the extension of dynamical mean-field
  theory\cite{Georges96} (DMFT) to
  nonequilibrium.%
  \cite{Freericks06,Eckstein08,Tsuji08,Eckstein08b,Tran08}
  DMFT, which is exact in the limit of infinite
  dimensions,\cite{Metzner89} can provide insights into the real-time
  evolution of strongly correlated systems in a nonperturbative way.
  
  The new one-particle description of TRPES given in
  Ref.~\onlinecite{Freericks08} leads to the question 
  whether real-time Green functions can be recovered fully from
  the time-dependent photoemission intensity, or whether parts of the
  electronic time evolution are not accessible by TRPES
  at all. Freericks~\ETAL{} discussed the case when electronic 
  {\em equilibrium states} are probed by pulses of finite time 
  duration.\cite{Freericks08,Freericks08b}
  This analysis covers experiments (e.g., those of 
  Ref.~\onlinecite{Perfetti06}) in which changes of external
  parameters such as the electronic temperature determine the dynamics
  of the electronic state, but the probe pulses are not short enough
  to resolve the thermalization of the electronic system in response
  to the pump pulse.  For this case the electronic state is characterized 
  by its frequency-dependent spectrum, and the photoemission intensity is 
  given by this spectrum, broadened in accordance with the frequency-time 
  uncertainty.\cite{Freericks08} Such a broadening can 
  hamper the determination of the electronic spectrum from photoemission 
  data; for the experiment of Ref.~\onlinecite{Perfetti06}, however, 
  it plays a minor role.\cite{Freericks08b}
  
  By contrast, in this paper we investigate TRPES with ultrashort 
  pulses that do resolve the thermalization of the electrons after 
  the pump pulse. We consider
  systems with a purely Hamiltonian time evolution involving only
  electronic degrees of freedom. The electronic state is then no 
  longer characterized only by a frequency-dependent spectrum, but 
  rather by real-time Green functions depending on two time variables.  
  We will show from the general theory of Ref.~\onlinecite{Freericks08} 
  that in this case the full time dependence on both time variables cannot 
  be recovered from the time-dependent photoemission intensity, no matter 
  how the pulse length of the probe pulse is chosen.  While time-resolved
  photoemission data can be predicted from calculated nonequilibrium
  Green functions,\cite{Freericks08} the converse procedure is thus
  impossible due to the frequency-time uncertainty relation. 
  We note that an analogous limitation is absent in
  time-resolved optical spectroscopy, where a two-time optical
  conductivity $\sigma(t,t')$ can be measured precisely
  by making the probe pulses sufficiently short. In nonequilibrium 
  DMFT, $\sigma(t,t')$ is directly related to momentum-averaged real-time 
  Green functions under certain conditions.\cite{Eckstein08b} 
  
  Below we employ the Falicov-Kimball model,\cite{Falicov69} which
  describes localized and mobile electrons on a lattice interacting
  via a local Hubbard interaction, to study the relation between
  nonequilibrium Green functions and time-resolved photoemission data
  in detail. We consider an idealized setup in which the system is
  suddenly driven out of a metallic or insulating equilibrium state,
  and subsequently relaxes to a new phase due to the Hamiltonian
  dynamics of the electrons. This model situation was recently solved
  with nonequilibrium DMFT.\cite{Eckstein08} We then study
  hypothetical time-resolved photoemission experiments during this
  relaxation process, and find that some aspects of the formation of
  the metallic state are indeed obscured in the photoemission spectrum
  due to the frequency-time uncertainty.  On the other hand, the
  relaxation of an excited Mott insulator leads to characteristic
  collapse-and-revival oscillations,\cite{Greiner02} which result in
  oscillating mid-gap weight in the time-resolved photoemission
  spectrum.  The above-mentioned uncertainty limitations
  notwithstanding, TRPES with ultrashort pulses is well-suited to
  characterize nonequilibrium states of correlated electron systems.
  However, it will often be necessary to analyze in detail how the
  time evolution of the Green function translates into the
  photoemission signal.

  The outline of the paper is as follows. In Sec.~\ref{sec:trpes} we
  briefly outline the microscopic formulation of TRPES derived in 
  Ref.~\onlinecite{Freericks08} and further discuss the role of 
  the frequency-time uncertainty in this theory. We then introduce
  the Falicov-Kimball model (Sec.~\ref{sec:fkm}) and discuss
  hypothetical time-resolved photoemission measurements on systems
  that relax to a metallic state (Sec.~\ref{sec:i2m}) and to an
  insulating state (Sec.~\ref{sec:m2i}). The discussion in
  Sec.~\ref{sec:conclusion} concludes the presentation.

\section{Time-resolved photoemission spectroscopy}
\label{sec:trpes}

  In photoemission experiments with both temporal and angular
  resolution the sample is probed with a finite pulse of definite
  wave vector\cite{footnote-deltaq} $\Vq$. The detector collects
  the photoelectrons which are emitted in a certain direction
  $\ke$, and it is sensitive to their kinetic energy $\ekin$ $=$
  $\hbar^2 k_e^2/2m$, but not to their arrival time ($\Vk_e$
  $=$ $\ke k_e$ is the photoelectron momentum). The time-resolved
  photoemission signal is thus proportional to the total number of
  electrons per solid angle $d\Omega_{\hat k_e}$ and energy
  interval $d\ekin$
  \begin{equation}
  \label{arpes-intensity}
  I(\ke,\ekin;\Vq,t_p) = \frac{dN(\ke,\ekin;\Vq,t_p)}
  {d\Omega_{\ke}d\ekin}
  \end{equation}
  that are emitted in response to a pulse that hits the sample at time
  $t_p$.\cite{Freericks08} This definition includes only 
  photoelectrons excited by the probe pulse and omits direct 
  photoemission due to the pump pulse.

  In Ref.~\onlinecite{Freericks08} an expression for the
  photoemission signal~(\ref{arpes-intensity}) was derived,
  using only the so-called sudden approximation,\cite{Hedin02}
  which neglects the interaction of photoelectrons with the remaining
  sample. The photoemission signal is then only related
  to matrix elements $M(\Vk,\Vq;\Vk_e)$ which couple Bloch states
  with quasi-momentum $\Vk$ in the solid and one-electron scattering
  states with asymptotic momentum $\Vk_e$ via absorption of a photon
  with momentum $\Vq$, and to the real-time one-particle Green function
  \begin{equation}
  \label{gkk}
  G^\les_{\Vk,\Vk'}(t,t')  = i \text{Tr}[\rho_0 c_{\Vk'}^\dagger(t')c_\Vk(t)].
  \end{equation}
  The latter incorporates the full nonequilibrium dynamics
  of the sample: $c_\Vk^{(\dagger)}(t)$ $=$ $U(t,\tmin)^\dagger c_\Vk^{(\dagger)} U(t,\tmin)$ 
  are annihilation (creation) operators for electrons in the solid with 
  momentum $\Vk$, whose propagation in time, with $U(t,\tmin)$ $=$ $T_\tau 
  \exp[-i\int_{\tmin}^{t} d\tau H(\tau)/\hbar]$, includes all external fields 
  except for the probe. The initial state at some early time $\tmin$
  is usually given by the thermal ensemble at temperature $T$, $\rho_0$ 
  $\propto$ $\exp[-H(\tmin)/T]$.
  The presence of the surface and the dependence of photoemission spectra
  on matrix elements can substantially complicate the comparison of 
  theoretical and experimental data for specific materials. In order 
  to reveal general aspects of TRPES 
  we thus resort to further approximations that are commonly made in this 
  context: (i) We assume that photoemission measures the bulk properties 
  of the sample which are contained in the momentum-diagonal Green 
  function $G^\les_{\Vk}(t,t')$ $\equiv$ $G^\les_{\Vk\Vk}(t,t')$ of the 
  infinite and translationally invariant system, and (ii), we take matrix 
  elements to be constant but satisfying momentum conservation in 
  the plane, $M(\Vq,\Vk;\Vk_e)$ $\equiv$ $M\delta_{\Vk_\pp+\Vq_\pp,\Vk_{e\pp}}$.
  The time-resolved  photoemission spectrum~(\ref{arpes-intensity}) is 
  then given by\cite{Freericks08}
  \begin{align}%
  \label{arpes-2d}%
  I(\ke,\ekin;\Vq,t_p)
  &\propto
  \sum_\Ks
  \delta_{\Vk_\pp+\Vq_\pp,\Vk_{e\pp}}
  \PES_\Ks(\ekin-cq-\Phi;t_p),
  \\
  \label{arpes-simple}
  \PES_\Ks(\omega;t_p)
  &= -i
  \int dt\!\! \int dt'
  S(t)S(t')
  \nonumber\\
  &\;\times \,\,\,
  e^{i\omega(t'-t)}
  G^\les_{\Ks}(t+t_p,t'+t_p),
  \end{align}%
  where $S(\tau)$ is the (real) pulse envelope function (centered 
  at $\tau$ $=$ $0$) and $\Phi$ is the work function of the solid. 
  Note that Eqn.~(\ref{arpes-2d}) and (\ref{arpes-simple})
  become exact for perfectly layered (two-dimensional) structures, 
  when $\Vk$ denotes the two-dimensional momentum of the sample. 
  In the following we will discuss the momentum- and
  frequency-dependent expression~(\ref{arpes-simple}). For simplicity
  we will refer to Eq.~(\ref{arpes-simple}) as the photoemission 
  intensity; observations made for this function presumably 
  persist after summation over some 
  part of the Brillouin zone [Eq.~(\ref{arpes-2d})].
  
  Eq.~(\ref{arpes-simple}) simplifies when the system is in equilibrium.
  In this case Green functions depend on the time difference only, and the
  Fourier transform is given by\cite{KadanoffBaym}
  \begin{align}
  g^\les_\Ks(\omega) &= \int\! dt\, e^{i\omega t} G^\les_\Ks(t,0)
  \label{gleseq}
  =
  2\pi i A_\Ks(\omega) f(\omega),
  \end{align}
  where $A_\Ks(\omega)$ is the equilibrium spectral 
  function,\cite{KadanoffBaym} and
  $f(\omega)$ $=$ $1/(e^{\omega/T}+1)$ is the Fermi function for
  temperature $T$. The photoemission intensity then reduces to
  \begin{equation}
  \label{pes-stationary}
  \PES_\Ks(\omega) =  \int d\omega' |\tilde S(\omega+\omega')|^2
   A_\Ks(\omega') f(\omega'),
  \end{equation}
  which is a convolution of the well-known expression
  $\PES_\Ks(\omega)$ $\propto$ $A_\Ks(\omega) f(\omega)$ for the
  intrinsic photocurrent in continuous beam experiments\cite{pes1,pes2}
  with the Fourier transform $\tilde S(\omega)$ $=$ 
  $\int\! dt\, S(t)
  e^{i\omega t}$ of the pulse envelope.  Due to the frequency-time
  uncertainty of the pulse, the frequency-dependent spectrum 
  is thus smoothened on a scale $\delta\omega$ $>$ $1/\delta t$ 
  when the pulse has a finite length $\delta t$, as
  discussed in Refs.~\onlinecite{Freericks08} and 
  \onlinecite{Freericks08b}.
  
  Here we study the case of an electronic system that is not in
  equilibrium.  The Green function $G^\les_\Ks(t,t+s)$ then contains
  important information both in the absolute time $t$ and in 
  the time difference $s$ between addition and removal of an 
  electron. However, when the probe pulse extends only over a finite 
  length $\delta$, the product $S(t)S(t')$ in Eq.~(\ref{arpes-simple}) 
  vanishes for all $t$ $-$ $t'$ $>$ $\delta$, and hence 
  $G^\les_\Ks(t,t+s)$ enters Eq.~(\ref{arpes-simple}) only 
  for $s$ $<$ $\delta$. It is therefore impossible to deduce 
  $G^\les_\Ks(t,t+s)$ from spectra that were recorded with pulses 
  of length $\delta$ $<$ $s$. In other words, the time resolution 
  (in~$t$) with which $G^\les_\Ks(t,t+s)$ can be measured is limited 
  by $s$. This also becomes clear when attempting to
  invert the convolution of $G^\les_\Ks(t,t')$ in 
  Eq.~(\ref{arpes-simple}): Starting from the Fourier transform
  \begin{align}
  \tilde \PES_\Ks(s;t_p) =
  \int d\omega e^{i\omega s}\,\PES_\Ks(\omega;t_p)\,,
  \label{its1}
  \end{align}
  and using, e.g., Gaussian pulses,
  \begin{align}
  \label{gauss}
  S(t) = \exp\Big(-\frac{t^2}{2\delta^2}\Big),
  \end{align}
  we obtain
  \begin{multline}
  \tilde \PES_\Ks(s;t_p)
  \propto
  \exp\Big(-\frac{s^2}{4\delta^2}\Big)
  \\
  \times
  \int \!dt\, G^\les_\Ks\Big(t_p+\frac{s}{2}+t,t_p-\frac{s}{2}+t\Big)
  \,\exp\Big(-\frac{t^2}{\delta^2}\Big).
  \label{its3}
  \end{multline}
  While the integral in~(\ref{its3}) apparently measures
  $G^\les_\Ks(t_p+s/2,t_p-s/2)$ with a time resolution of $\delta$, it
  is practically impossible to choose $\delta$ $\ll$ $s$ because then
  the result vanishes compared to any noise added to $\tilde
  \PES_\Ks(s;t_p)$, due to the Gaussian prefactor (whose form is due
  to (\ref{gauss}) but the suppression of the signal for $\delta$
  $\ll$ $s$ is independent of the pulse shape).  We conclude that the
  nonequilibrium two-time Green function cannot be fully measured by
  means of TRPES, and one must always carefully analyze how the time
  evolution of the Green function translates into the photoemission
  signal for a given theoretical model.  This will be
  illustrated for the Falicov-Kimball model below.

  \section{The Falicov-Kimball model in nonequilibrium}
  \label{sec:fkm}

  In the remaining part of this paper we concentrate on one specific
  model for electronic dynamics in a single band, the Falicov-Kimball
  model.\cite{Falicov69} This lattice model describes
  itinerant~($\flavorc$) and immobile~($\flavorf$) electrons
  which interact via the local Coulomb repulsion~$U$. The Hamiltonian is
  given by
  \begin{align}%
  \label{FKM}
  H = \sum_{ij} V_{ij} c_{i\flavorc}^{\mydagger} c_{j\flavorc}^{\phdagger}
  + U \sum_{i} n_{i\flavorc} n_{i\flavorf}
  -\sum_{i\sigma}\mu_\sigma n_{i\sigma},
  \end{align}%
  where $c_{i\sigma}^{(\mydagger)}$ are annihilation (creation)
  operators for the two species of fermions on lattice site $i$,
  and $n_{i\sigma}$ $=$
  $c_{i\sigma}^{\mydagger} c_{i\sigma}^{\phdagger}$ is the
  corresponding density ($\sigma$ $=$ $\flavorc$, $\flavorf$).
  Hopping between sites $i$ and $j$ (with amplitude $V_{ij}$) is
  possible only for the mobile~($\flavorc$) particles.
  The Falicov-Kimball model has been an important benchmark
  for the development of DMFT in equilibrium, because the effective
  single-site problem for the mobile particles is quadratic and can
  be solved exactly.\cite{Brandt89} This model currently plays a similar role
  for nonequilibrium DMFT,\cite{Freericks06,Eckstein08,Tsuji08,Eckstein08b,Tran08}
  in particular since no appropriate real-time impurity solver is
  yet available for the Hubbard model.

  In spite of its apparent simplicity the Falicov-Kimball model has a
  rich equilibrium phase diagram containing metallic, insulating, and
  charge-ordered phases.\cite{Freericks03} In the following we
  consider only the homogeneous phase at half-filling for both
  particle species ($n_\flavorc$ $=$ $n_\flavorf$ $=$ $1/2$), which in
  equilibrium undergoes a metal-insulator transition at a critical
  interaction $U$ $=$ $U_c$ on the order of the 
  bandwidth.\cite{Brandt89,Freericks03,vanDongen90,vanDongen92}
  This phase is studied in an idealized nonequilibrium situation, by
  preparing the system in thermal equilibrium for times $t$ $<$ $0$,
  and changing the interaction parameter $U$ abruptly at $t$ $=$
  $0$.\cite{Eckstein08} Of course within a more realistic 
  description of the pump pulse the system would not be excited 
  into a state that is an equilibrium state of any simple 
  Hamiltonian. Nonetheless, the {\em interaction quench} we 
  study here allows us to discuss time-resolved photoemission 
  signals for a situation in which the electronic dynamics drives 
  the system between different phases. Since the exact DMFT solution for
  the interaction quench is available\cite{Eckstein08} we can directly
  relate Green functions and photoemisson signals. In
  particular, we will focus on two specific phenomena, namely (i) the
  formation of narrow quasiparticle resonances during the buildup of
  the metallic state (Sec.~\ref{sec:i2m}), and (ii) coherent
  collective oscillations after an excitation of the insulating phase
  (Sec.~\ref{sec:m2i}).

  In the following we analyze the time-resolved photoemission signal
  of the mobile~($\flavorc$) electrons using Gaussian pulses (\ref{gauss})
  and omit the time-independent contribution of the immobile ($\flavorf$)
  electrons. As in Ref.~\onlinecite{Eckstein08} we assume a
  semielliptic density of states $\rho(\epsilon)$ $=$ 
  $\sqrt{4V^2-\epsilon^2}/(2\pi V^2)$ with half-bandwidth $2V$ for the
  single-particle energies $\epsilon_\Vk$, which are the eigenvalues
  of the hopping matrix $V_{ij}$.  Exact expressions for the real-time
  Green functions $G^\les_{\Vk}(t,t')$ of the mobile electrons are
  given in the Appendix (from now on we suppress the index $\flavorc$
  of the mobile electrons).  We take $V$ $=$ $1$ as the unit of energy,
  so that the full bandwidth is $4$ and the critical interaction is 
  given by $U_c$ $=$ $2V$ =
  $2$.\cite{vanDongen92} We will also set $\hbar$ $=$ $1$, setting the
  unit of time as $\hbar/V$. For example, for $V$ $=$ 1~eV we have
  $\hbar/V$ $=$ 0.66~fs.

  \section{Pumping the insulator into a metallic state}
  \label{sec:i2m}

  \begin{figure}
   \centerline{\includegraphics[width=\hsize]{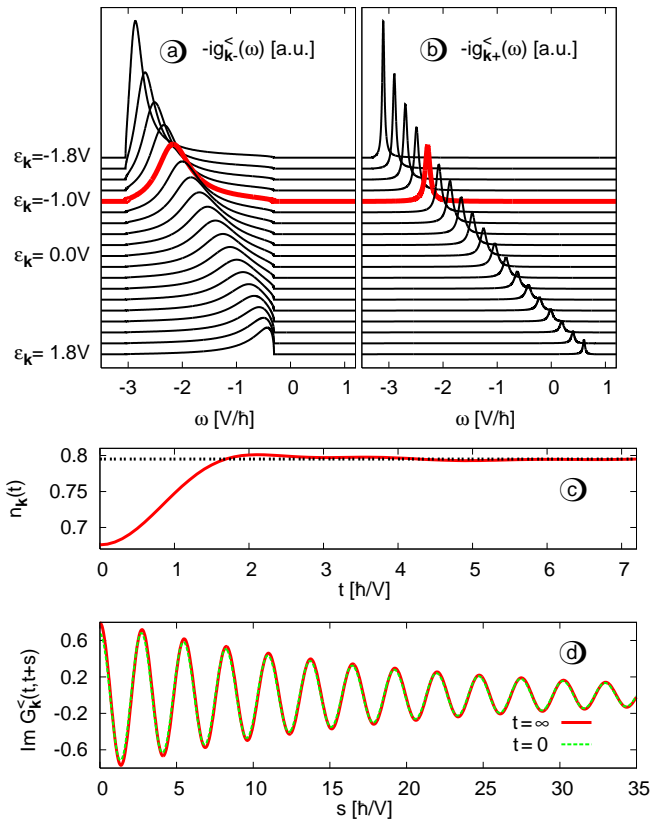}}
   \caption{The momentum-dependent Green function $g^\les_{\Vk\pm}(\omega)$
   [Eq.~(\ref{gkinf})] in the initial (a) and final (b) state for
   the quench from $U$ $=$ $3$ to $U$ $=$ $0.5$ ($n_c$ $=$ $n_f$ $=$
   $1/2$; temperature $T$ $=$ $0$). Due to the
   local self-energy in DMFT, $G^\les_\Vk(t,t')$
   depends on $\Vk$ only via $\epsilon_\Vk$ in the homogeneous
   phase. (c) Density $n_\Vk(t)$ for momentum $\Vk$ with
   $\epsilon_\Vk$ $=$ $-1$ [thick red curve in (a) and (b)].
   The horizontal line is at $n_\Vk(\infty)$.
   (d) Green function $G^\les_\Vk(t+s,t)$ for the same
   $\epsilon_\Vk$. Differences between the Green functions for
   $t$ $=$ $0$ and $t$ $=$ $\infty$ are best visible around
   $s$ $=$ $0$; their decay is almost identical.}
   \label{fig:gU3-0.5}
   \end{figure}

   In this section we investigate the formation of a metallic state in
   real time, similar to pump-induced insulator-to-metal transitions
   on ultrashort time scales.%
   \cite{Ogasawara00,Iwai03,Okamoto07,Perfetti06} In the
   Falicov-Kimball model, such a process takes place after a quench
   from an insulating state to the metallic parameter regime, which we
   consider now.  We prepare the initial state at $U$ $=$ $3$ and
   temperature $T$ $=$ $0$, and change the interaction abruptly to $U$
   $=$ $0.5$ at time $t$ $=$ $0$.  Subsequently the system relaxes to
   a new stationary state, in which Green functions depend on time
   difference only.\cite{Eckstein08} In the following we first discuss the
   real-time Green functions for this process and then the
   corresponding time-resolved photoemission signal.

  \subsection{Real-time Green functions}
  
  The difference between the initial and final state is evident from the
  momentum-diagonal Green function $G^\les_{\Vk}(t,t')$ of the mobile
  particles [Eq.~(\ref{gkk})] and its Fourier transform
  \begin{align}%
  \label{gkinf}
  g^\les_{\Vk\mp}(\omega)
  \equiv
  \lim_{t \to \mp \infty}
  \int \!ds \,e^{i\omega s}\,
  G^\les_{\Vk}(t+s,t)
  \end{align}%
  in the limit $t$ $=$ $\mp\infty$, respectively: While 
  $g^\les_{\Vk-}(\omega)$ has a
  broad maximum (Fig.~\ref{fig:gU3-0.5}a), a sharp peak in
  $g^\les_{\Vk+}(\omega)$ indicates that quasiparticle excitations
  have a long lifetime in the final state (Fig.~\ref{fig:gU3-0.5}b).
  Note that for a quench in the Falicov-Kimball model the final
  state always retains memory on the initial
  configuration.\cite{Eckstein08} This memory is
  contained in a nonuniversal occupation function $F(\omega)$ which
  replaces the Fermi function $f(\omega)$ in Eq.~(\ref{gleseq}), i.e.,
  $g^\les_{\Vk+}(\omega)$ $=$ $2\pi i A_\Vk(\omega) F(\omega)$ (see
  Appendix).
  
  The development of the metallic state with its sharp
  quasiparticle-like resonances can be observed from the full time
  dependence of the momentum-diagonal Green function
  $G^\les_\Vk(t,t')$. In particular, we characterize the transition by
  means of (i) the total spectral weight, i.e., the momentum
  occupation $n_\Vk(t)$ $=$ $-iG^\les_\Vk(t,t)$, and (ii) the decay of
  $G^\les_\Vk(t,t+s)$ as a function of $s$ $>$ $0$. From the
  latter one can read off the lifetime of a hole which is 
  created at time $t$ in the nonequilibrium state.  The time 
  evolution of these two quantities is similar for all $\Vk$; 
  it is shown for one representative value of $\Vk$ in 
  Fig.~\ref{fig:gU3-0.5}c and~\ref{fig:gU3-0.5}d (namely 
  $\epsilon_\Vk$ $=$ $-1$, marked by the thick red 
  line in Fig.~\ref{fig:gU3-0.5}a and \ref{fig:gU3-0.5}b): (i) Relaxation of 
  the momentum occupation $n_\Vk(t)$ takes place on a
  time scale on the order of the inverse bandwidth; after this very
  short time interval the final value $n_\Vk(\infty)$ is almost
  reached, and a slower relaxation follows (Fig.~\ref{fig:gU3-0.5}c).
  A similar behavior was observed previously for the time dependence
  of the number of doubly occupied sites.\cite{Eckstein08} 
  (ii) As a function of $s$, the Green function $G^\les_\Vk(t,t+s)$ 
  decays slow compared to the inverse bandwidth (Fig.~\ref{fig:gU3-0.5}d). 
  For $t$ $=$ $\infty$, this is in accordance with the sharp peak in
  $g^\les_{\Vk+}(\omega)$. However, the fast decay is observed
  even for holes that are created at $t$ $=$ $0^+$, when the system 
  is still in the insulating state, indicating that the 
  lifetime of hole excitations depends only weakly on the 
  time of their creation.  The reason for this
  behavior is that in the Falicov-Kimball model scattering occurs only
  between mobile and immobile particles and is thus determined by the
  Hamiltonian and the (initial) configuration of the immobile
  particles. If scattering occurred between two mobile electron
  species (e.g., as in the Hubbard model), then we would expect the
  shape of the quasiparticle resonances to depend 
  on the quantum state of the mobile
  particles as well, and therefore would expect it to change
  considerably during the relaxation process.
  
  \begin{figure}
   \centerline{\includegraphics[width=\hsize]{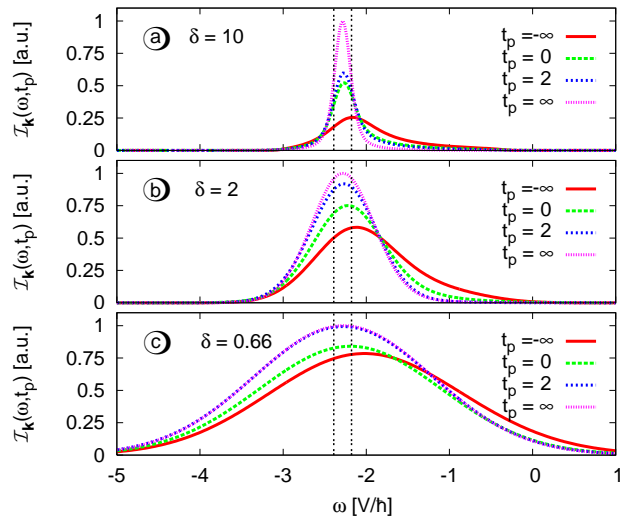}}
   \caption{Photoemission signal [Eq.~(\ref{arpes-simple})] (a.u.) for the same situation
   as the Green functions in Fig.~\ref{fig:gU3-0.5} ($\epsilon_\Vk$ $=$ $-1$),
   using Gaussian probe  envelopes~(\ref{gauss}) with $\delta$ $=$ $10$ (a), 
   $\delta$ $=$ $2$ (b),  and $\delta$ $=$ $0.66$ (c). The vertical 
   dashed lines are explained  in the text. $t_p$ and $\delta$ are in units 
   of $\hbar/V$ $=$ $1$.}
   \label{fig:pesU3-0.5}
  \end{figure}

  \subsection{Photoemission spectrum}

  As discussed in the introduction, it is a central question whether
  the time-resolved photoemission spectrum $\PES_{\Vk}(\omega;t_p)$
  [Eq.~(\ref{arpes-simple})] contains the same information as the
  Green function $G^\les_\Vk(t,t')$.  For the present case, the answer 
  is no: In the last subsection we saw that holes decay 
  at a rate $\Gamma$ $\ll$ $V/\hbar$, which depends only weakly
  on the time $t$ when the hole is created, even for
  short times $0$ $<$ $t$ $<$ $\hbar/V$. To establish this 
  behavior from the photoemission intensity, however, one would
  have to measure $G_\Vk(t,t+1/\Gamma)$ with time resolution better
  than~$\hbar/V$, which is impossible according to the discussion at
  the end of Sec.~\ref{sec:trpes}.
  
  The effect of the frequency-time uncertainty can be seen in detail
  in the redistribution of spectral weight in the photoemission signal
  $\PES_\Vk(\omega;t_p)$ as a function of the probe time~$t_p$ (for
  $\epsilon_\Vk$ $=$ $-1$, Fig.~\ref{fig:pesU3-0.5}). When
  the system is probed in a stationary state ($t_p$ $=$ $\pm \infty$),
  the intensity is given by $g^\les_{\Vk\pm}(\omega)$ folded with the
  spectrum $|\tilde S(\omega)|^2$ $=$
  $2\pi\delta^2\exp(-\delta^2\omega^2)$ of the probe pulse [cf.
  Eq.~(\ref{pes-stationary})]. This broadening completely washes out
  the peak for short pulses ($\delta$ $=$ $0.66$ in units of
  $\hbar/V$, Fig.~\ref{fig:pesU3-0.5}c). On the other hand, pulses
  much longer than the inverse bandwidth do not resolve the fast
  relaxation which is essentially complete for $t$ $>$ $2$, but rather
  show a time dependence even for $t$ $\gtrsim$ 2 ($\delta$ $=$ $10$,
  Fig.~\ref{fig:pesU3-0.5}a). For intermediate pulse length neither
  time nor frequency dependence is resolved ($\delta$ $=$ $2$,
  Fig.~\ref{fig:pesU3-0.5}b).

\begin{figure}
   \centerline{\includegraphics[width=\hsize]{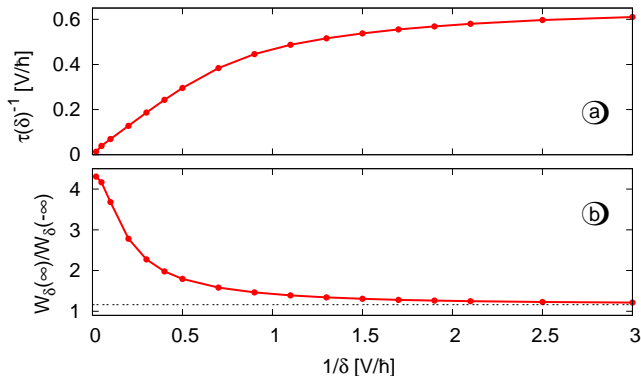}}
   \caption{(a) Relaxation time $\tau(\delta)$ [Eq.~(\ref{taueta}),
   $\eta$ $=$ $0.95$] for the spectral width in the central peak
   region of the photoemission spectrum at $\epsilon_\Vk$ $=$ $-1$
   (region between the vertical dashed lines in Fig.~\ref{fig:pesU3-0.5}),
   plotted against the pulse length $\delta$. 
   (b) Ratio $W_\delta(\infty)/W_\delta(-\infty)$. The
   horizontal dashed line is at $n_\Vk(\infty)/n_\Vk(-\infty)$.
   }
   \label{fig:pestransfer}
  \end{figure}

  For a quantitative analysis we now consider the weight in the
  central region of the peak in the photoemission spectrum,
  \begin{align}%
  \label{integrated-pes}
  W_\delta(t_p) \equiv \int_a^b d\omega\,\PES_\Vk(\omega;t_p).
  \end{align}%
  In the present case we use $a$ $=$ $-2.39$ and $b$ $=$ $-2.17$ as
  indicated by the vertical dashed lines in Fig.~\ref{fig:pesU3-0.5};
  they enclose twice the full width at half maximum of a Lorentzian fit to
  the peak in $A_\Vk(\omega)$. We then define a time $\tau(\delta)$ at
  which the relaxation of $W_\delta(t_p)$ is essentially complete,
  \begin{align}%
  \label{taueta}
  \frac{W_\delta[\tau(\delta)]-W_\delta(-\infty)}
  {W_\delta(\infty)-W_\delta(-\infty)}
  =
  \eta,
  \end{align}%
  with $\eta$ close to one; we use $\eta$ $=$ 0.95.  In
  Fig.~\ref{fig:pestransfer}a we see first of all that $\tau(\delta)$
  is proportional to the pulse length for large $\delta$, which is due
  to the fact that long pulses merely average the spectrum of the
  final and initial state.  Hence $\tau(\delta)$ does not yield any
  information about intrinsic relaxation times for $\delta$ $\gtrsim$
  $2$.  On the other hand, for $\delta$ $\lesssim$ $2$ the ratio
  $W_\delta(\infty)/W_\delta(-\infty)$ approaches the value
  $n_\Vk(\infty)/n_\Vk(-\infty)$ that one would obtain by integrating
  over the whole spectrum instead of the peak region alone
  (Fig.~\ref{fig:pestransfer}b); this is due to the insufficient
  energy resolution.  Therefore only the relaxation time of the whole
  spectral width, i.e., of the momentum occupation $n_\Vk(t)$, can be
  determined reliably.

  \section{Oscillations of an excited Mott insulator}
  \label{sec:m2i}
  
  In this section we investigate the relaxation dynamics of a Mott
  insulator in which a metallic state has been created by the pump
  pulse. In the Falicov-Kimball model this can be simulated in the
  strongly interacting regime ($U$ $=$ $10$) by preparing the system
  in a metallic state ($U$ $=$ $1$, temperature $T$ $=$ $0$) at $t$
  $=$ $0$. Again we first discuss the real-time Green functions for
  this situation and then the time-resolved photoemission signal that
  corresponds to them.

  \subsection{Real-time Green functions}

\begin{figure}
   \centerline{\includegraphics[width=\hsize]{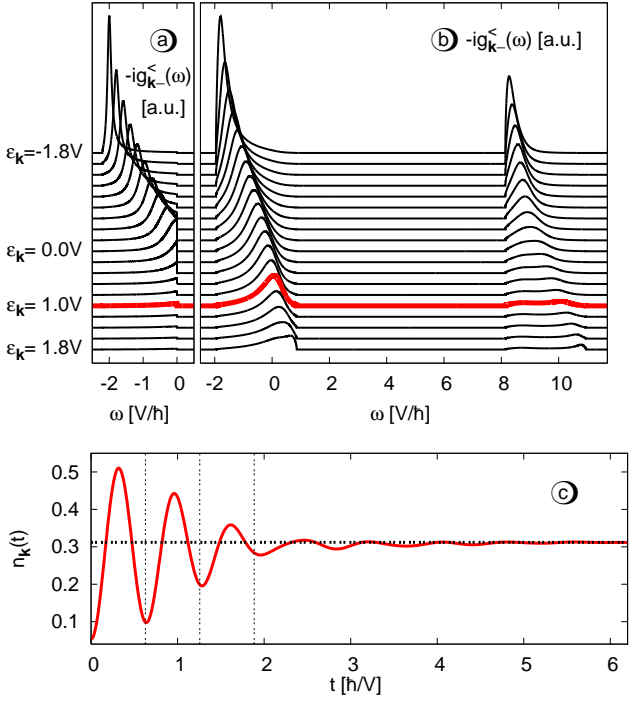}}
   \caption{The momentum-dependent Green function $g^\les_{\Vk\pm}(\omega)$
   [Eq.~(\ref{gkinf})] in the initial (a) and final (b) state for
   the quench from $U$ $=$ $1$ to $U$ $=$ $10$ ($n_c$ $=$ $n_f$ $=$
   $1/2$; temperature $T$ $=$ $0$). (c) Density $n_\Vk(t)$
   for momentum $\Vk$ with  $\epsilon_\Vk$ $=$ $1$ [thick red
   curve in (a) and (b)]. The vertical dashed lines are at multiples
   of the fundamental oscillation period $2\pi/U$. The
   horizontal line is $n_\Vk(\infty)$.}
   \label{fig:gkU1-10}
   \end{figure}

  In the present case the momentum-dependent Green function evolves
  from a well-defined quasiparticle band at $t$ $=$ $-\infty$, which
  is cut off by the Fermi function (Fig.~\ref{fig:gkU1-10}a), to a
  gapped spectrum at $t$ $=$ $\infty$ (Fig.~\ref{fig:gkU1-10}b). Note
  that in the final state spectral weight remains in the upper band
  because the system is strongly excited with respect to the
  insulating ground state at $U$ $=$ $10$, and there is no coupling to
  an environment to which this excess energy could be passed during
  the relaxation process.
  
   As in the previous section we consider a representative momentum
   ($\epsilon_\Vk$  $=$ $1$), chosen such that $n_\Vk$ is
   small in the metallic state and increases after the quench
   (Fig.~\ref{fig:gkU1-10}c).  Again this relaxation takes place on a
   very short time scale (on the order of the inverse bandwidth), but
   now $n_\Vk$ passes through a series of damped oscillations with
   period $2\pi/U$ before reaching its final value.  These
   oscillations are characteristic for the dynamics of a Mott
   insulator which is dominated by a Hubbard-type density interaction
   $U \sum_i n_{i\flavorf} n_{i\flavorc}$. In fact, if  the
   Hamiltonian were given  only by this interaction term, then the time
   evolution operator $\exp(-itU \sum_i n_{i\flavorf} n_{i\flavorc})$
   would itself be $2\pi/U$-periodic,\cite{Greiner02} and hence
   oscillations would occur in all nonlocal quantities. These so-called
   collapse-and-revival oscillations were first observed and described
   in experiments with ultracold atomic gases,\cite{Greiner02} where
   the Hamiltonian of the system can be designed in a controlled way.
   We will now discuss the fingerprint of these oscillations in the
   time-resolved photoemission spectrum.

  \subsection{Photoemission spectrum}

\begin{figure}
  \centerline{\includegraphics[width=0.9\hsize]{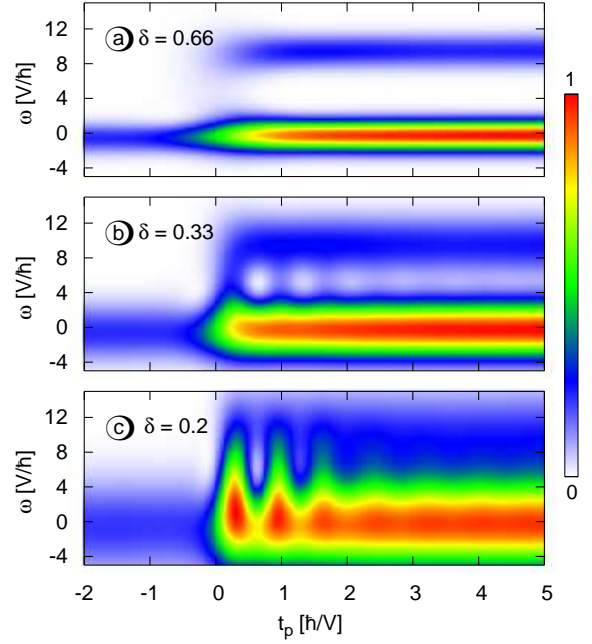}}
  \caption{Quench from $U$ $=$ $1$ to $U$ $=$ $10$: Photoemission signal (a.u.)
  [Eq.~(\ref{arpes-simple})] for $\epsilon_\Vk$ $=$ $1$, and Gaussian probe
  envelopes~(\ref{gauss}) with $\delta$ $=$ $0.66$ (a), $\delta$ $=$
  $0.33$ (b), and $\delta$ $=$ $0.2$ (c).
  Pulse lengths
  $\delta$ are in units of $\hbar/V$ $=$ $1$.}
  \label{fig:pesU1-10}
  \end{figure}

  In. Fig.~\ref{fig:pesU1-10} the angular-resolved photoemission
  spectrum $\PES_{\Vk}(\omega;t_p)$ is plotted for the same fixed
  momentum [$\epsilon_\Vk$ $=$ $1$, using Gaussian
  pulses~(\ref{gauss})]. All features of the spectrum except for its
  total weight, which is proportional to $n_\Vk(t)$, are washed out
  for short pulses ($\delta$ $=$ $0.2$, Fig.~\ref{fig:pesU1-10}c),
  whereas long pulses show the formation of a gap, but
  cannot resolve the oscillating nature of the
  state ($\delta$ $=$ $0.66$, Fig.~\ref{fig:pesU1-10}a).  For
  intermediate pulses, however, both the $2\pi/U$-periodicity and the
  gap become visible ($\delta$ $=$ $0.33$, Fig.~\ref{fig:pesU1-10}b).

  Interestingly, the coherent oscillations are most pronounced in the
  center of the gap (Fig.~\ref{fig:pesU1-10}b). This observation
  can be understood from the atomic limit of the Hamiltonian~(\ref{FKM}),
  i.e., for $V_{ij}$ $=$ 0.  In the strongly interacting regime $U$
  $\gg$ $V$ the atomic limit gives a good description of the transient
  behavior at short times $t$ $\lesssim$ $\hbar/V$.  For the
  interaction term alone, $H_U$ $=$ $U \sum_i n_{i\flavorf}
  n_{i\flavorc}$ $-$ $\sum_{i\sigma} \mu_\sigma n_{i\sigma}$ the time
  evolution of annihilation operators 
  is given by $e^{i H_U t} c_{j\sigma}e^{-i H_U t}$ $=$
  $e^{i\mu_\sigma t}[c_{j\sigma}$ $+$
  $(e^{-itU}-1)c_{j\sigma}n_{i\bar\sigma}]$.  For $t$,$t'$ $>$ $0$,
  the Green function then follows as
  \begin{subequations}%
  \begin{multline}%
  G^\les_{\Vk\sigma}(t,t') =
  ie^{i\mu_\sigma(t-t')}
  [A_{\Vk\sigma} +
  B_{\Vk\sigma}e^{it'U}
  +
  \\
  B_{\Vk\sigma}^* e^{-itU}
  +
  C_{\Vk\sigma}e^{iU(t'-t)}],
  \end{multline}%
  with
  \begin{align}%
  A_{\Vk\sigma}
  &= \sum_{ij} e^{i\Vk(\VR_i-\VR_j)}
  \expval{(1-n_{j\bar\sigma})c_{j\sigma}^{\mydagger} c_{i\sigma}^{\phdagger}(1-n_{i\bar\sigma})}_0
  \\
  B_{\Vk\sigma}
  &= \sum_{ij} e^{i\Vk(\VR_i-\VR_j)}
  \expval{n_{j\bar\sigma}c_{j\sigma}^{\mydagger} c_{i\sigma}^{\phdagger}}_0
  \\
  C_{\Vk\sigma}
  &= \sum_{ij} e^{i\Vk(\VR_i-\VR_j)}
  \expval{n_{j\bar\sigma}c_{j\sigma}^{\mydagger} c_{i\sigma}^{\phdagger}n_{i\bar\sigma}}_0,
  \end{align}%
  \end{subequations}%
  and $\expval{\cdot}_0$ is the expectation
  value in the (arbitrary) state at $t$ $=$ $0$ immediately after the
  pump. Inserting this expression into Eq.~(\ref{arpes-simple}) we
  find, for $t_p$ $\gg$ $\delta$,
  \begin{multline}%
  \PES_\Ks(\omega;t_p)
  \,\,\propto\,\,
  A_{\Vk\sigma} |\tilde S(\omega+\mu_\sigma)|^2
  +
  C_{\Vk\sigma} |\tilde S(\omega+\mu_\sigma-U)|^2
  \\
  +\;2\,\text{Re}[
  \tilde S(\omega)\tilde S(\omega+\mu_\sigma-U) B_{\Vk\sigma} e^{it_pU}
  ].
  \end{multline}%
  The first two terms are centered in the upper and lower Hubbard
  bands at $\omega$ $=$ $-\mu_\sigma$ and $\omega$ $=$ $U-\mu_\sigma$
  and do not change with time. The third term, which oscillates with
  period $2\pi/U$, has its maximum where $\tilde S(\omega+\mu_\sigma)$
  and $\tilde S(\omega+\mu_\sigma-U)$ overlap.  For Gaussian pulses,
  this is precisely the center of the gap because $\tilde S(\omega)
  \tilde S(\omega-U)$ $\propto$ $e^{-(\omega-U/2)^2/2\delta^2}
  e^{-U^2/\delta^2}$.
  
  This discussion shows that on short time scales the observed time-dependent 
  spectrum in the Falicov-Kimball model for large interactions 
  resembled that of the atomic limit.  Note that the initial state
  at $t$ $=$ $0$ determines only the weight of the three components,
  but not the frequency of the oscillations. The oscillating midgap
  weight is thus a universal property of the Mott insulator, which is
  largely independent of the excitation process.  Via this universal
  feature it may eventually become possible to observe
  collapse-and-revival oscillations in TRPES experiments on correlated
  materials.

  \section{Conclusion}
  \label{sec:conclusion}

  In this paper we analyzed hypothetical time-resolved photoemission
  experiments on correlated electron systems that are not in
  equilibrium, building on the general theory of
  Ref.~\onlinecite{Freericks08}. We showed that the two-time Green
  function, which characterizes the nonequilibrium state of the
  electrons, cannot be fully measured with TRPES, no matter how long
  or short the probe pulses are chosen.  For example, using DMFT for
  the Falicov-Kimball model we found that in the buildup of the
  metallic state the Green function of the transient state cannot be
  determined from the photoemission signal.  On the other hand, if an
  excited Mott insulator is created by the pump pulse, its
  characteristic collapse-and-revival oscillations can nevertheless be
  inferred because they correspond to oscillating weight in the center
  of the Mott gap in the time-dependent photoemission spectrum.
  
  TRPES is in some sense complementary to time-resolved optical
  spectroscopy, which measures the two-time optical conductivity
  $\sigma(t,t')$. Under certain conditions, the latter is obtained in
  DMFT from a momentum-averaged product of two Green functions that
  also enter the expression for the photoemission spectrum.\cite{Eckstein08b} The
  dependence of $\sigma(t,t')$ on both $t$ and $t'$ can be measured
  precisely with sufficiently short probe pulses, unaffected by any
  minimum uncertainty, but unlike in photoemission spectroscopy there
  is no sensitivity toward specific momenta~$\bm{k}$.
 
  In conclusion, we showed that in spite of the frequency-time
  uncertainty of the probe pulse, TRPES has the potential to discover
  fascinating details of the electronic thermalization process.
  Unlike for conventional photoemission on systems in equilibrium,
  however, the time-dependent photoemission signal does not yield the
  real-time Green function directly, so that more detailed comparisons
  to theoretical predictions are needed.


  \section*{Acknowledgements}
  We thank Dieter Vollhardt for valuable discussions.
  M.E.\ acknowledges support by Studienstiftung des Deutschen Volkes.
  This work was supported in part by the SFB 484 of the Deutsche
  Forschungsgemeinschaft.


  \appendix

  \section{Calculation of DMFT Green functions}
  \label{app:green}

  \subsection{Contour Green functions}

  In this appendix, which is a direct extension of the work
  presented in Ref.~\onlinecite{Eckstein08}, we give the detailed
  derivation of the real-time Green functions for the interaction 
  quench in the Falicov-Kimball model, using DMFT for nonequilibrium.%
  \cite{Freericks06}. 
  
  The retarded, advanced, and lesser Green functions are defined by
  \begin{subequations}%
  \label{realtimeg}
  \begin{align}%
  G^\ret_\Ks(t,t') &=
  -i\Theta(t-t') \text{Tr}[\rho_0\{
  c^{\phdagger}_\Ks(t),c^{\mydagger}_\Ks(t')\}]
  \label{gret}
  \\
  G^\adv_\Ks(t,t') &=
  i\Theta(t'-t) \text{Tr}[\rho_0\{
  c^{\phdagger}_\Ks(t),c^{\mydagger}_\Ks(t')\}]
  \label{gadv}
  \\G^\les_\Ks(t,t') &=
  \;i \text{Tr}[\rho_0 c^{\mydagger}_\Ks(t')c^{\phdagger}_\Ks(t)]
  \label{glesser}
  \end{align}%
  \end{subequations}%
  respectively, where $c_\Ks(t)$ and $\rho_0$ are defined below
  Eq.~(\ref{gkk}).  DMFT for nonequilibrium is based on the Keldysh
  formalism,\cite{Keldysh64,Keldyshintro} which yields the
  contour-ordered Green function $G_\Ks(t,t')$ $=$ $-i\expval{\TC
  c_\Ks^\phdagger(t)c_\Ks^\mydagger(t')}$ with time
  arguments on the contour $\CC$ that runs from $\tmin$ to $\tmax$ on
  the real axis, then from $\tmax$ to $\tmin$, and finally to
  $\tmin-i\beta$ on the imaginary time axis ($\beta$: inverse
  temperature).  Retarded, advanced, and lesser Green functions
  (\ref{realtimeg}) are obtained from the real-time components of the
  contour Green function,\cite{Keldyshintro}
  \begin{subequations}%
  \label{contourrelations}
  \begin{align}%
  G_\Ks^\ret(t,t')
  &=
  [G_\Ks^{11}(t,t')-G_\Ks^{12}(t,t')]
  \\
  G_\Ks^\adv(t,t')
  &=
  [G_\Ks^{11}(t,t')-G_\Ks^{21}(t,t')]
  \\
  G_\Ks^\les(t,t')
  &=
  G_\Ks^{12}(t,t'),
  \end{align}%
  \end{subequations}%
  where superscripts refer to the two time arguments: $1$, $2$, and
  $3$ indicates whether a time argument is on the upper, lower, or
  vertical part of the contour, respectively.  These relations and
  also the symmetries
  \begin{subequations}%
  \label{contoursymmetry}
  \begin{align}%
  G_\Ks^\ret(t,t')
  &=
  G_\Ks^\adv(t',t)^*
  \\
  G_\Ks^\les(t,t')
  &=
  -G_\Ks^\les(t',t)^*.
  \end{align}%
  \end{subequations}%
  hold for all contour Green functions considered here.  Furthermore,
  the contour Green functions obey antiperiodic boundary conditions in
  both contour arguments:
  \begin{subequations}%
  \label{boundaryconditions}
  \begin{align}%
  G_\Ks^{\nu1}(t,\tmin)
  &=
  -
  G_\Ks^{\nu3}(t,\tmin\!-\!i\beta),
  \\
  G_\Ks^{1\nu}(\tmin,t')
  &=
  -
  G_\Ks^{3\nu}(\tmin\!-\!i\beta,t'),
  \end{align}%
  \end{subequations}%
  for $\nu$ $=$ $1,2,3$.
  
  For the lattice Hamiltonian (\ref{FKM}), the interacting contour 
  Green function satisfies a Dyson 
  equation\cite{Keldyshintro}
  \begin{subequations}%
  \label{dysonk}
  \begin{align}%
  [(\cG^{-1}_\Ks - \Sigma_\Ks) \convz
  G_\Ks](t,t') = \delC(t,t'),
  \end{align}%
  where $\Sigma_\Ks(t,t')$ is the contour self-energy, and
  $\cG_\Ks(t,t')$ is the noninteracting Green function,
  whose inverse
  \begin{align}%
  \cG^{-1}_\Ks(t,t') = \delC(t,t')[(i\DC{t} + \mu_\sigma)
  - \epsilon_\Ks(t)]
  \end{align}%
  \end{subequations}%
  can be written as differential operator on the contour.
  Here we assumed a homogeneous state, and $\epsilon_\Ks$
  $=$ $\sum_j V_{ij}^\sigma \exp[i\Vk(\VR_j-\VR_i)]$ are the
  single-particle band energies. We also introduced 
  the convolution $(f \myast g)(t,t')$ $=$ 
  $\int_\CC d\bar{t}f(t,\bar{t})g(\bar{t},t')$ of two functions
  along the contour, the contour delta function $\delC\!(t,t')$
  [defined by $\int_\CC \!d\bar{t}\,f(\bar{t})\delC(\bar{t},t)$
  $=$ $f(t)$], and the contour derivative 
  $\partial_t^\CC$.\cite{Freericks06}
  Taking into account the boundary 
  conditions~(\ref{boundaryconditions}), the 
  integro-differential equation (\ref{dysonk}) has a unique 
  solution for $G_\Ks(t,t')$.
  
  \subsection{DMFT equations}
  
  In DMFT the self-energy is assumed to be local in space, and hence
  independent of $\Vk$ for a homogeneous state. This approximation is
  exact in the limit of infinite spatial dimensions,\cite{Metzner89}
  both for equilibrium and for the Keldysh self-energy.\cite{Freericks06}
  The local self-energy $\Sigma_\sigma(t,t')$ and the local Green function
  $G_{\sigma}(t,t')$ $\equiv$
  $G_{ii\sigma}(t,t')$ $=$
  $\sum_\Vk G_\Ks(t,t')$
  are then calculated from an auxiliary problem in which the
  degrees of freedom at a single lattice site $i$ are coupled
  to some unknown environment, which must be determined
  self-consistently. For the Falicov-Kimball model, the auxiliary 
  problem is quadratic\cite{Brandt89,Freericks06} such that the 
  equations of motion can be solved explicitly.

  The DMFT equations for the interaction quench in the Falicov-Kimball
  model were derived in Ref.~\onlinecite{Eckstein08}. 
  In the following we state these equations without derivation, and then give 
  details of the solution. In particular we calculate the momentum-dependent 
  Green function $G_\Vk(t,t')$ $\equiv$ $G_{\Vk\flavorc}(t,t')$ of the mobile~($\flavorc$)
  particles, which is needed for the photoemission intensity 
  (\ref{arpes-simple}). From now on we consider only properties of 
  the mobile particles, and omit the index $\flavorc$.
  
  The local Green function of the mobile particles is given by the sum
   \begin{subequations}%
    \label{g-from-lambda}%
    \begin{align}%
      G(t,t') 
      &=
      w_0 \Dloc(t,t')
      +
      w_1 \Bloc(t,t')
      \,,\label{g=b+d}
    \end{align}%
    of the local Green functions $\Dloc(t,t')$ and $\Bloc(t,t')$
    at sites with zero and one immobile~($\flavorf$)  particle, respectively, 
    weighted with the average density $w_1$ $=$ $1-w_0$ of immobile 
    particles. The functions $\Dloc(t,t')$ and $\Bloc(t,t')$ obey 
    the equations of motion
    \begin{align}%
      &{}[i\partial_{t}^\CC\! + \mu]
      \Dloc(t,t')
      -
      (\Lambda \convz \Dloc)(t,t')
      =
      \delC\!(t,t')
      \,,\label{eq:eqm-d}
      \\
      &{}[i\partial_{t}^\CC\! + \mu - U(t)]
      \Bloc(t,t')
      -
      (\Lambda \convz \Bloc)(t,t')
      =
      \delC\!(t,t')
      ,\label{eq:eqm-b}
    \end{align}%
  \end{subequations}%
  and boundary conditions~(\ref{boundaryconditions}). For a quench
  the interaction is piecewise constant in time, $U(t)$ 
  $=$ $\Theta(t) U_+$ $+$ $\Theta(-t)U_-$. The effective medium
  propagator $\Lambda(t,t')$ must be determined self-consistently. 
  For a semielliptic density of states of the mobile particles,
  which we adopt in the following, the self-consistency cycle can 
  be condensed into closed form,\cite{Eckstein08}
  \begin{align}%
  \label{sce}
  \Lambda(t,t') = V^2 G(t,t'),
  \end{align}%
  where $2V$ is the half bandwidth of the density of states.
  Eqn. (\ref{g-from-lambda}) and (\ref{sce}) form a complete set of 
  equations for the local Green function. The local self-energy 
  $\Sigma$ of the mobile particles is then obtained  from the Dyson 
  equation of the local problem, $[(i\DC{t} + \mu)G(t,t')$ $-$
  $[(\Lambda +\Sigma)\convz G](t,t')$ $=$ $\delC(t,t')$.
  Together with Eq.~(\ref{g-from-lambda}), this is easily transformed 
  into
  \begin{align}%
  \label{eqm-sigma}
  w_1U(t)\Bloc(t,t') = (\Sigma \convz G)(t,t').
  \end{align}%
  Finally the local self-energy
  $\Sigma(t,t')$ is inserted into the lattice Dyson equation 
  (\ref{dysonk}),
  \begin{align}%
  \label{dysondmft}
  (i\DC{t} + \mu
  - \epsilon_\Vk)G_\Vk(t,t')-
  [\Sigma \convz G_\Vk](t,t') = \delC(t,t'),
  \end{align}%
  which yields the $\Vk$-dependent Green functions.
  Note that $G_\Vk(t,t')$ depends on momentum only via the single 
  particle energy $\epsilon_\Vk$, because we assumed a homogeneous state.
  
  \subsection{Langreth rules}
  
  To solve the contour equations (\ref{eq:eqm-d}), (\ref{eq:eqm-b}),
  (\ref{eqm-sigma}), and (\ref{dysondmft}) we first rewrite them in
  terms of their retarded and lesser components, using the identities
  (\ref{contourrelations}). In effect, this means that contour
  derivatives $C(t,t')$ $=$ $\partial_t^\CC A(t,t')$ are replaced by
  real-time derivatives,\cite{Keldyshintro}
  \begin{subequations}%
  \begin{align}%
  \label{langreth-deriv-ret}
  C^\ret(t,t') &= \partial_t A^\ret(t,t') \\
  \label{langreth-deriv-les}
  C^\les(t,t') &= \partial_t A^\les(t,t'),
  \end{align}%
  \end{subequations}%
  and convolutions $C(t,t')$ $=$ $[A\convz B](t,t')$ of two contour
  Green functions $A$ and $B$ are expressed in terms of their
  retarded, advanced, and lesser components according to the Langreth
  rules\cite{Keldyshintro}
  \begin{subequations}%
  \label{langreth}
  \begin{align}%
  \label{langreth-ret}
  C^\ret(t,t')
  &=
  \int_{t'}^{t} d\tb\, A^\ret(t,\tb) B^\ret(\tb,t');
  \\
  \label{langreth-lesser}
  C^\les(t,t')
  \!
  &=
  \!\!
  \int_{-\infty}^{t'}
  d\tb\,
  A^\les(t,\tb) B^\adv(\tb,t')
  \nonumber\\
  &\hspace*{12mm}+
  \int_{-\infty}^{t} d\tb\,
  A^\ret(t,\tb) B^\les(\tb,t').
  \end{align}%
  \end{subequations}%
  The integral boundaries account for the fact that retarded
  (advanced) Green functions $A^{\ret(\adv)}(t,t')$ vanish when $t$
  $<$ $t'$ ($t$ $>$ $t'$).  Furthermore, we shifted $\tmin$ $\to$
  $-\infty$ in the second equation, such that the convolution extends
  over the whole axis but contributions from the vertical part at
  $\tmin-i\tau$ can be dropped. This step is discussed in further
  detail below.
  
  The contour delta function on the right-hand-side of
  (\ref{eq:eqm-d}), (\ref{eq:eqm-b}), and (\ref{dysondmft}) vanishes
  when the lesser component is taken, and it is replaced by the usual
  delta function $\delta(t-t')$ for the retarded components. However,
  because any retarded function $A^\ret(t,t')$ vanishes for $t$ $<$
  $t'$, retarded equations of motion are only considered for $t$ $>$
  $t'$, and the initial value at $t$ $=$ $t'$ is determined by the 
  weight of the delta function and the derivative operator. In 
  particular, we obtain
  \begin{align}%
  \label{ic-ret}
  G^\ret_\Vk(t,t) = \Bloc^\ret(t,t) = 
  \Dloc^\ret(t,t) =-i
  \end{align}%
  from Eqn.~(\ref{eq:eqm-d}), (\ref{eq:eqm-b}), and (\ref{dysondmft}).
  These conditions follow also directly from the anticommutation 
  relation of creation and annihilation operators.
  
  \subsection{Stationary states}
  
  For the interaction quench we treat the equations of motion
  separately in the four regions where both $t$ and $t'$ do not change
  sign; we introduce additional subscripts $+$ and $-$ which indicate
  whether the time arguments are greater or less than zero,
  respectively. Inserting (\ref{langreth-deriv-ret}) and
  (\ref{langreth-ret}) into Eq.~(\ref{g-from-lambda}) yields a closed
  set of equations for $\Bloc^\ret(t,t')$ and $\Dloc^\ret(t,t')$,
  \begin{subequations}%
  \label{eqm-ret}
  \begin{align}%
  \label{sce-ret}
  \Lambda^\ret(t,t')
  =
  V^2
  &[w_1 \Bloc^\ret(t,t')
  +
  w_0 \Dloc^\ret(t,t')]
  \\
  [i\partial_t  + \mu
  &] \Dloc^\ret(t,t')
  =
  \int_{t'}^{t} \!d\tb\,
  \Lambda^\ret(t,\tb) \Dloc^\ret(\tb,t')
  \\
  [i\partial_t  + \mu- U(t)
  &] \Bloc^\ret(t,t')
  =
  \int_{t'}^{t} \!d\tb\,
  \Lambda^\ret(t,\tb) \Bloc^\ret(\tb,t'),
  \end{align}%
  \end{subequations}%
  which must be solved for $t$ $>$ $t'$, using the initial condition
  (\ref{ic-ret}). The self-consistency equation (\ref{sce}) was used
  in Eq.~(\ref{sce-ret}).  Note that in Eq.~(\ref{eqm-ret}), Green
  functions with both time arguments greater or less that zero, i.e.,
  the ($++$) and ($--$) components, do not mix with other components.
  Because $U(t)$ is constant for $t$ $>$ $0$ and $t$ $<$ $0$,
  respectively, the solutions of (\ref{eqm-ret}) are thus
  translationally invariant in time when both $t$ and $t'$ have the same
  sign, and we make the ansatz
  \begin{subequations}%
  \begin{align}%
  A^\ret_{\pm\pm}(t,t')
  &=
  a^\ret_\pm(t-t')
  \\
  \label{ftreteq}
  \ft a^\ret_\pm(z) 
  &=
  \int_0^\infty \!\!\!ds\, e^{izs}\, a^\ret_\pm(s)
  \end{align}%
  \end{subequations}%
  for all contour functions $A$ $=$ $G$, $\Bloc$, $\Dloc$, $\Lambda$,
  $G_\Vk$, and $\Sigma$ (with $a$ $=$ $g$, $\bloc$, $\dloc$,
  $\lambda$, $g_\Vk$, and $\sigma$, respectively). Using this ansatz
  in Eq.~(\ref{eqm-ret}) we obtain a set of cubic equations, 
  \begin{subequations}%
    \label{eq:cubic}%
    \begin{align}%
      \ft g^\ret_\pm(z) &= w_0\ft \dloc^\ret_\pm(z) + w_1\ft \dloc^\ret_\pm(z)
      \,,
      \\
      \ft \dloc^\ret_\pm(z) &= [z+\mu - V^2 \ft g^\ret_\pm(z)]^{-1}
      \,
      \\
      \ft \bloc^\ret_\pm(z) &= [z+\mu - V^2 \ft g^\ret_\pm(z) - U_\pm ]^{-1}
      \,.
    \end{align}%
  \end{subequations}%
  that can be solved analytically. These cubic equations are well-known
  from the DMFT solution of the Falicov-Kimball model 
  in equilibrium.\cite{vanDongen90} This is of course expected 
  when both $t$ and $t'$ $<$ $0$, because before the quench the 
  system indeed is in an equilibrium state. In a similar way, 
  the retarded ($++$) and 
  ($--$) components of $\Sigma$ and $G_\Vk$ are obtained from 
  Eq.~(\ref{eqm-sigma}) and (\ref{dysondmft}), 
  \begin{align}%
  \ft \sigma^\ret_\pm(z) 
  &= 
  w_1 U_\pm \ft \bloc^\ret_\pm(z)/\ft g^\ret_\pm(z)
  \\
  \ft g^\ret_{\Vk\pm}(z) 
  &= 
  [z+\mu-\epsilon_\Vk-\ft \sigma^\ret_\pm(z)]^{-1}.
  \end{align}%
  Furthermore, advanced Green functions are directly related to the
  retarded ones by symmetry (\ref{contoursymmetry}), so that we have
  \begin{subequations}%
  \begin{align}%
  A^\adv_{\pm\pm}(t,t')
  &=
  a^\adv_\pm(t-t')
  \\
  \label{ftadveq}
  \ft a^\adv_\pm(z) 
  &=
  \int_{-\infty}^0 \!\!\!ds\, e^{izs}\, a^\adv_\pm(s)
  = \ft a^\ret_\pm(z^*)^*.
  \end{align}%
  \end{subequations}%

  The lesser Green functions are translationally invariant in time only
  for both $t$ and $t'$ $<$ $0$ [($--$) component], when the system is
  still in equilibrium. One then has\cite{Keldyshintro}
  \begin{subequations}%
  \label{lessereq}
  \begin{align}%
  A^\les_\ee(t,t')
  &=
  \int \frac{d\omega}{2\pi}\,e^{i\omega(t'-t)}
  \ft a^\les_\e(\omega),
  \\
  \label{leseqw}
  \ft a^\les_\e(\omega) 
  &= 
  f(\omega) [\ft a^\adv_\e(\omega) -\ft a^\ret_\e(\omega)],
  \end{align}%
  \end{subequations}%
  where $f(\omega)$ $=$ $1/(e^{\omega/T}+1)$ is the Fermi 
  function, and $\ft a^\adv_\e(\omega)$ $-$ $\ft a^\ret_\e(\omega)$
  $=$ $-2i\,\text{Im}\,a^\ret_\e(\omega)$ is proportional to
  the spectrum of the equilibrium Green function. Mathematically 
  this follows from the
  solutions of the equations of motion on the full contour, 
  including the vertical part, and taking into account the 
  antiperiodic boundary conditions. For the quench we use 
  (\ref{lessereq}) as {\em initial condition} for the lesser 
  components; only then can we then let $\tmin$ $\to$ $-\infty$, and
  disregard the vertical part in the Langreth 
  rule (\ref{langreth-lesser}).
  
  On the other hand, we show below that in the limit where both $t$
  and $t'$ tend to $\infty$ (but their difference is finite), the
  lesser ($++$) components take a form very similar to
  (\ref{lessereq}),
  \begin{subequations}%
  \label{lesserfi}
  \begin{align}%
  \label{lesfitt}
  \lim_{t\to\infty} A^\les_\zz(t+s,t)
  &=
  \int \frac{d\omega}{2\pi}\,e^{-i\omega s}
  \ft a^\les_\z(\omega),
  \\
  \label{lesfiw}
  \ft a^\les_\z(\omega) 
  &= 
  F(\omega) [\ft a^\adv_\z(\omega) -\ft a^\ret_\z(\omega)].
  \end{align}%
  \end{subequations}%
  The function $F(\omega)$ is common for all $a$ $=$ $g$, $\bloc$,
  $\dloc$, $\sigma$, and $g_\Vk$. One can in fact directly see from
  the equations of motion (\ref{eq:eqm-d}), (\ref{eq:eqm-b}),
  (\ref{eqm-sigma}), and (\ref{dysondmft}) for the lesser component
  that {\em if} the stationary limit (\ref{lesfitt}) exists, {\em
    then} Green functions $\ft a^\les_+(\omega)$ must have this common
  factor $F(\omega)$. To find this factor, however, the equations of
  motion must be solved, because it contains the entire information
  about the initial state.

  \subsection{Double Fourier transforms}

  We now consider the cases with one or two positive time arguments,
  i.e., after the quench.  We introduce double Fourier transforms
  \begin{subequations}%
  \label{dft-ret}
  \begin{align}%
  &\ft A^\ret_\ze (z,\eta)
  =
  \int_0^\infty \!\!dt\, e^{izt}
  \int_{-\infty}^0 \!\!dt'\, e^{i\eta t'}
  A^\ret_\ze (t,t')
  \\
  &\ft A^\adv_\ez (\eta,z)
  =
  \int_0^\infty \!\!dt\, e^{izt}
  \int_{-\infty}^0 \!\!dt'\, e^{i\eta t'}
  A^\adv_\ez (t',t),
  \end{align}%
  \end{subequations}%
  for retarded and advanced components,
  \begin{subequations}%
  \label{dft-les+-}
  \begin{align}%
  &\int \frac{d\omega}{2\pi}\,e^{-i\omega t'}
  \ft A^\les_\ze(z,\omega)
  =
  \int_0^\infty  \!\!dt\, e^{izt}
  A^\les_\ze(t,t')
  \\
  &\int \frac{d\omega}{2\pi}\,e^{-i\omega t'}
  \ft A^\les_\ez(\omega,z)
  =
  \int_0^\infty  \!\!dt\, e^{iz t}
  A^\les_\ez(t',t)
  \end{align}%
  \end{subequations}%
  for the lesser components with mixed time arguments
  (which holds for $t'$ $<$ $0$), and
  \begin{align}%
  \label{dft-les++}
  &\ft A^\les_\zz (z,\eta)
  =
  \int_0^\infty \!\!dt\, e^{izt}
  \int_0^\infty \!\!dt'\, e^{i\eta t'}
  A^\les_\zz (t,t')
  \end{align}%
  for the lesser Green function with both time arguments
  after the quench. In this subsection we derive
  explicit expressions for $\ft A^\ret_\ze(z,\omega)$, 
  $\ft A^\les_\ze(z,\omega)$, and $\ft A^\les_\zz(z,\omega)$;
  the remaining are then obtained by symmetry (\ref{contoursymmetry}),
  \begin{subequations}%
  \begin{align}%
  \ft A^\adv_\ez(\eta,z) 
  &=
  \ft A^\ret_\ze(-\eta^*,-z^*)^*
  \\
  \ft A^\les_\ez(\omega,z)
  &=
  - \ft A^\les_\ze(-z^*,-\omega)^*.
  \end{align}%
  \end{subequations}%
  Using Langreth rules (\ref{langreth}) once again yields
  for convolutions $C$ $=$ $A\convz B$,
  \begin{widetext}%
  \begin{subequations}%
  \label{langreth-long}
  \begin{align}%
  \label{langreth-long-ret+-}
  \ft C^\ret_\ze (z,-\omega)
  &=
  \ft A^\ret_\ze(z,-\omega) \ft b^\ret_\e(\omega) +
  \ft a^\ret_\z(z) \ft B^\ret_\ze(z,-\omega)
  \\
  \label{langreth-long-less+-}
  \ft C^\les_\ze (z,-\omega)
  &=
  \ft A^\les_\ze(z,-\omega) \ft b^\adv_\e(\omega)
  +
  \ft a^\ret_\z(z) \ft B^\les_\ze(z,-\omega)
  +
  \ft A^\ret_\ze(z,-\omega) \ft b^\les_\e(\omega)
  \\
  \label{langreth-long-less++}
  \ft C^\les_\zz (z,\eta)
  &=
  \ft A^\les_\zz(z,\eta)\ft b^\adv_\z(-\eta)
  +
  \ft a^\ret_\z(z) \ft B^\les_\zz(z,\eta)
  +
  \int \frac{d\omega}{2\pi}\,
  [
  \ft A^\les_\ze(z,-\omega) \ft B^\adv_\ez(\omega,\eta)
  +
  \ft A^\ret_\ze(z,-\omega) \ft B^\les_\ez(\omega,\eta)
  ].
  \end{align}%
  \end{subequations}%
  \end{widetext}
  Furthermore, the derivative  $C$ $=$ $\partial_t^\CC A(t,t')$
  translates into
  \begin{subequations}%
  \label{langreth-deriv}
  \begin{align}%
  \label{langreth-deriv-ret+-}
  \ft C^\ret_\ze(z,\eta)
  &=
  z \ft A^\ret(z,\eta)
  -
  i \ft a^\ret_\e(-\eta)
  \\
  \label{langreth-deriv-less+-}
  \ft C^\les_\ze(z,\omega)
  &=
  z \ft A^\les_\ze(z,-\omega)
  -
  i \ft a^\les_\e(\omega)
  \\
  \label{langreth-deriv-less++}
  \ft C^\les_\zz(z,\eta)
  &=
  z \ft A^\les_\zz(z,\eta)
  -
  i \int \frac{d\omega}{2\pi}\, \ft A^\les_\ez(\omega,\eta),
  \end{align}%
  \end{subequations}%
  where one must use the continuity of the components
  at the boundary $t$ $=$ $0$ and $t'$ $=$ $0$, e.g.,
  $A^\ret_\ze(0,t')$ $=$ $A^\ret_\ee(0,t')$.

  Using (\ref{langreth-long}) and (\ref{langreth-deriv}) one can
  rewrite Eq.~(\ref{g-from-lambda}) for the various components and
  solve them using the self-consistency equation~(\ref{sce}). For
  instance, we obtain the ($+-$) component of the retarded Green
  functions as
  \begin{align}%
  \ft \Bloc^\ret_\ze(z,\eta)
  &=
  [\ft \Lambda^\ret_\ze(z,\eta)+i]\bloc^\ret_\z(z)\bloc^\ret_\e(-\eta),
  \\
  \ft \Dloc^\ret_\ze(z,\eta)
  &=
  [\ft \Lambda^\ret_\ze(z,\eta)+i]\dloc^\ret_\z(z)\dloc^\ret_\e(-\eta),  
  \end{align}%
  where Eq.~(\ref{eq:cubic}) was used once. Together with the
  self-consistency~(\ref{sce}), this is a simple linear equation for
  $\Lambda^\ret_\ze(z,\eta)$. In a similar way all components are
  determined successively: Starting from the retarded ($+-$) 
  and advanced ($-+$) components, the results enter the lesser ($+-$)
  and lesser ($-+$) components [cf. Eqn.~(\ref{langreth-long-less+-})
  and (\ref{langreth-deriv-less+-})], which in turn enter the
  equations for the lesser ($++$) component [cf.
  Eqn.~(\ref{langreth-long-less++}) and
  (\ref{langreth-deriv-less++})]. The procedure is repeated for
  Eq.~(\ref{g-from-lambda}), Eq.~(\ref{eqm-sigma}), and finally for
  the lattice Dyson equation (\ref{dysondmft}), which yields the
  momentum-dependent Green function $G_\Vk(t,t')$.

  For completeness we state the final result for $G$, $\Bloc$, 
  and $G_\Vk$. For this we introduce the abbreviations
  \begin{subequations}%
  \label{kernelm}
  \begin{align}%
  M^{xy}_{\alpha\beta}
  &=[1-V^2(
  w_1 \ft \bloc^x_\alpha\ft\bloc^y_\beta
  +
  w_0 \ft\dloc^x_\alpha\ft\dloc^y_\beta
  )]^{-1},
  \\
  \kappa^x_\alpha
  &= \ft\bloc^x_\alpha \ft\dloc^x_\alpha / \ft g^x_\alpha.
  \\
  K^{xy}_{\alpha\beta}
  &=1+
  w_1w_0U_+U_-\kappa^x_\alpha \kappa^y_\beta M^{xy}_{\alpha\beta},
  \end{align}%
  \end{subequations}%
  with superscripts $x$,$y$ $\in$ $\{\ret,\adv\}$, and subscripts
  $\alpha$, $\beta$ $\in$ $\{+,-\}$, and we use the convention
  that the variables of a function $a^x_\alpha$ is
  (i) $z$ when $x$ $=$ $\ret$ and $\alpha$ $=$ $+$,
  (ii) $-\eta$ when $x$ $=$ $\adv$ and $\alpha$ $=$ $+$,
  and (iii) $\omega$ when $\alpha$ $=$ $-$.
  In terms of these expressions the final
  result is
  \begin{subequations}%
  \begin{align}%
  \ft G^\ret_\ze(z,-\omega) 
  &= 
  iV^{-2}(M^{\ret\ret}_\ze-1),
  \\
  \ft \Bloc^\ret_\ze(z,-\omega) 
  &= 
  i\ft \bloc^\ret_\z \ft \bloc^\ret_\e M^{\ret\ret}_\ze,
  \\
  \ft G^\ret_{\Vk\ze}(z,-\omega)
  &=
  i\ft g^\ret_{\Vk\z} \ft g^\ret_{\Vk\e}
  K^{\ret\ret}_\ze,
  \end{align}%
  \end{subequations}%
  and
  \begin{subequations}%
  \begin{align}%
  \ft G^\les_\ze(z,-\omega) 
  &= 
  if(\omega)(
  M^{\ret\adv}_\ze-M^{\ret\ret}_\ze)/V^2,
  \\
  \ft \Bloc^\les_\ze(z,-\omega) 
  &=
  if(\omega)\ft \bloc^\ret_\z(
  \ft \bloc^\adv_\e M^{\ret\adv}_\ze-
  \ft \bloc^\ret_\e M^{\ret\ret}_\ze
  ),
  \\
  \ft G^\les_{\Vk\ze}(z,-\omega) 
  &=
  if(\omega)\ft g^\ret_{\Vk\z}(
  \ft g^\adv_{\Vk\e} K^{\ret\adv}_\ze-
  \ft g^\ret_{\Vk\e} K^{\ret\ret}_\ze
  ).
  \end{align}%
  \end{subequations}%
  The lesser ($++$) component can be written in the form
  \begin{subequations}%
  \label{lesser++final}%
  \begin{align}%
  \label{lesser++general}%
  A^\les_\zz(z,\eta) =
  \ft F(z,\eta) \frac{\ft a^\adv_\z-\ft a^\ret_\z}{\eta + z} + F_A(z,\eta)
  \end{align}%
  for $A$ $=$ $G$, $\Bloc$, and $G_\Vk$,
  where
  \begin{align}%
  \ft F(z,\eta)
  &=
  \int \frac{d\omega}{2\pi} f(\omega)
  \frac{
  M^{\ret\ret}_\ze
  +
  M^{\adv\adv}_\ze
  -
  M^{\ret\adv}_\ze
  -
  M^{\adv\ret}_\ze
  }
  {z+\eta +V^2(\ft g^\adv_\z- \ft g^\ret_\z)},
  \end{align}%
  and
  \begin{align}%
  F_G(z,\eta)
  &=
  \ft F(z,\eta)/V^2,
  \\
  F_\Bloc(z,\eta)
  &=
  \ft \bloc^\ret_\z
  \ft \bloc^\adv_\z
  \int \frac{d\omega}{2\pi} f(\omega)
  \nonumber\\&\;\times
  (
  \ft \bloc^\adv_\e
  M^{\adv\adv}_\ze
  M^{\ret\adv}_\ze
  -
  \ft \bloc^\ret_\e
  M^{\ret\ret}_\ze
  M^{\adv\ret}_\ze
  ),
  \end{align}%
  \begin{multline}%
  F_{G_\Vk}(z,\eta)
  =
  \ft g^\ret_{\Vk\z} 
  \ft g^\adv_{\Vk\z} 
  \Big\{ -\ft F(z,\eta) + 
  \int \frac{d\omega}{2\pi} f(\omega)
  \\\times
  {\Big[}
  w_0 w_1 U_+^2
  \ft \kappa^\ret_\z
  \ft \kappa^\adv_\z
  (
  \ft \kappa^\adv_\e
  M^{\adv\adv}_\ze
  M^{\ret\adv}_\ze
  -
  \ft \kappa^\ret_\e
  M^{\ret\ret}_\ze
  M^{\adv\ret}_\ze
  )
  \\
  +\ft g^\adv_{\Vk\e}
  K^{\adv\adv}_\ze
  K^{\ret\adv}_\ze
  -
  \ft g^\ret_{\Vk\e}
  K^{\ret\ret}_\ze
  K^{\adv\ret}_\ze
  \Big]\Big\}.
  \end{multline}%
  \end{subequations}%

  \subsection{Back transformation}

  To obtain the physical real-time Green functions, 
  we have to invert the double Fourier transformations
  (\ref{dft-ret}), (\ref{dft-les+-}), and (\ref{dft-les++}),
  using the final expressions (\ref{kernelm}) through 
  (\ref{lesser++final}). Here we give an explicit formula 
  for the partially Fourier-transformed lesser component
  \begin{align}%
  \label{gles-wt}
  \ft A^\les(\omega,t) = \int ds \, e^{i \omega s} A^\les(t+s,t),
  \end{align}%
  ($A$ $=$ $G$, $\Bloc$, and $G_\Vk$). The singularity at 
  $\eta+z$ $=$ $0$ in Eq.~(\ref{lesser++general}) 
  determines (\ref{gles-wt}) in the limit 
  $t$ $\to$ $\infty$,
  \begin{align}%
  \ft a^\les_\z(\omega)
  &\equiv
  \lim_{t\to\infty}
  \ft A^\les(\omega,t) 
  \nonumber\\&
  =
  \text{Im} \ft F(\omega,-\omega) 
  [\ft a^\adv_\z(\omega) - \ft a^\ret_\z(\omega)],
  \end{align}%
  which is of the form discussed above [cf. Eq.~(\ref{lesserfi})],
  with $F(\omega)$ $=$ $\text{Im} \ft F(\omega,-\omega) $.
  The full result is given by
  \begin{subequations}%
  \begin{multline}%
  \ft A^\les(\omega,t) =
  \sum_\pm \Theta(\pm t)
  \\\times
  \left[
  \ft a^\les_\pm(\omega) + 
  e^{-i\omega t}
  \int \frac{d\eta}{2\pi}\, e^{-it\eta}\hat A_\pm(\omega,\eta)
  \right],
  \end{multline}%
  where
  \begin{align}%
  \hat A_-(\omega,\eta)
  &=
  \ft A^\les_\ze(\omega,\eta)
  +
  \frac{\ft a^\les_-(-\eta)-\ft a^\les_-(\omega)}{i(\eta+\omega)},
  \end{align}%
  \begin{multline}%
  \label{lasteqn}
  \hat A_+(\omega,\eta)
  =
  2 i \text{Im}
  \frac{\ft F(\omega,\eta)[\ft a^\adv_\z(-\eta)-\ft a^\ret_\z(\omega)]}
  {\omega+\eta}
  \\
  \,\,+\,
   F_A(\omega,\eta)
   +
  \int \frac{d\omega'}{2\pi i}
  \frac{\ft A^\les_\ez(\omega',\eta)}{\omega-i0-\omega'}\,,
  \end{multline}%
  \end{subequations}%
  and the components for $A$ $=$ $G$, $\Bloc$, and $G_\Vk$ 
  were given in the previous subsection.
  Note that the first term on the right hand side of 
  Eq.~(\ref{lasteqn}) is regular at $\eta$ $=$ $-\omega$, 
  because both $\ft F(\omega,\eta)$ and $\ft a^\adv_\z(-\eta)$ $-$ 
  $\ft a^\ret_\z(\omega)]$ are then purely imaginary.


\begin{thebibliography}{99}
\bibitem{footnote-deltaq}
    A short pulse has necessarily finite extent in space, but its 
    spread in wavenumber can nevertheless be neglected due to the 
    steep dispersion of light: even when the frequency width of the 
    pulse is on the order of a typical bandwidth ($\hbar\delta\omega$ $=$ $1\,eV$), 
    the spread in wavenumber, $\delta q$ $=$ $|\delta\omega/c|$ $\approx$ $0.0005/$\AA,
    which is small compared to the typical extent of the Brillouin zone.
\end{thebibliography}
\end{document}